\input psfig.sty
\documentstyle{cupconf}

\ifoldfss
\else
  \ifnfssone
    \newmathalphabet{\mathit}
      \addtoversion{normal}{\mathit}{cmr}{m}{it}
      \addtoversion{bold}{\mathit}{cmr}{bx}{it}
    \newmathalphabet{\mathcal}
      \addtoversion{normal}{\mathcal}{cmsy}{m}{n}
    \else
    \ifnfsstwo
    \fi
  \fi
\fi

\def\hexnumber#1{\ifcase#1 0\or1\or2\or3\or4\or5\or6\or7\or8\or9\or
 A\or B\or C\or D\or E\or F\fi }

\makeatletter
\ifx\CUP@mtlplain@loaded\undefined
\else
\fi
\makeatother
 \makeatletter
 \ifx\CUP@mtlplain@loaded\undefined
   \font\tenbmi=cmmib10 at 10pt
   \font\sevenbmi=cmmib10 at 7pt
   \font\fivebmi=cmmib10 at 5pt

   \newfam\bmifam
   \textfont\bmifam=\tenbmi
   \scriptfont\bmifam=\sevenbmi
   \scriptscriptfont\bmifam=\fivebmi
   
 \fi
 \makeatother
\ifnfsstwo

\fi
\ifnfssone

\fi
\ifoldfss

\fi

\mathchardef\varLambda="0103

\makeatletter
\ifx\CUP@mtlplain@loaded\undefined
\else
\fi
\makeatother
\makeatletter
\ifx\CUP@mtlplain@loaded\undefined
  \font\tenbms=cmbsy10
  \font\sevenbms=cmbsy10 at 7pt
  \font\fivebms=cmbsy10 at 5pt
  \newfam\bmsfam
  \textfont\bmsfam=\tenbms
  \scriptfont\bmsfam=\sevenbms
  \scriptscriptfont\bmsfam=\fivebms

  \edef\bsy@{\hexnumber\bmsfam}
  \mathchardef\bnabla="0\bsy@72
\fi
\makeatother
\title[Polarimetric VLBI observations of 0735+178 ]{Polarimetric VLBI observations of 0735+178}

\author[I. Agudo {\it et al.\/}]%
{I\ls V\ls \'A\ls N\ns A\ls G\ls U\ls D\ls O$^1$,\ns
J.\ls \ls L.\ns G\ls \'O\ls M\ls E\ls Z$^1$,\ns 
D.\ls C.\ns G\ls A\ls B\ls U\ls Z\ls D\ls A$^2$,\ns\\ 
J.\ls \ls C.\ns G\ls U\ls I\ls R\ls A\ls D\ls O$^3$,\ns 
A.\ls A\ls L\ls B\ls E\ls  R\ls D\ls I$^1$,\ns
A.\ls P.\ns M\ls A\ls R\ls S\ls C\ls H\ls E\ls R$^4$,\ns\\ 
M.\ls \ls A.\ns A\ls L\ls O\ls Y$^5$,\ns  
\and \ns J.\ns \ls M.\ns M\ls A\ls R\ls T\ls \'I\ls$^3$}

\affiliation{$^1$Instituto de Astrof\'{\i}sica de Andaluc\'{\i}a, CSIC,Apartado 3004, 18080 Granada, Spain\\[\affilskip]
$^2$Joint Institute for VLBI in Europe, Postbus 2, 7990 AA Dwingeloo, The Netherlands\\[\affilskip]
$^3$Departamento de Astronom\'{\i}a y Astrof\'{\i}sica, Universidad de Valencia, 46100 Burjassot (Valencia), Spain\\[\affilskip]
$^4$Institute for Astrophysical Research, Boston University, 725 Commonwealth Avenue, Boston, MA 02215, USA\\[\affilskip]
$^5$Max-Planck-Institut f\"ur Astrophysik, Karl-Schwarzschild-Str.
1, D-85748 Garching, Germany}

\begin{document}
\ifnfssone
\else
  \ifnfsstwo
  \else
    \ifoldfss
      \let\mathcal\cal
      \let\mathrm\rm
      \let\mathsf\sf
    \fi
  \fi
\fi

\maketitle

\begin{abstract}

  We present a new centimeter polarimetric VLBI image of the BL~Lac
object 0735+178. This source exhibits one of the most pronounced curvatures
observed in jets of AGNs, with two sharp apparent bends of 90 degrees within
the inner 2 milliarcseconds from the core. Through the analysis of the data
gathered over the past decades we study whether this curvature is produced 
by precession of the jet with ejection of ballistic components, or a 
precession in such a way that components' velocity vector are always parallel 
to the jet axis. These possibilities are also studied by comparison with 3D 
hydrodynamic relativistic simulations of precessing jets.

\end{abstract}

\firstsection

\section{Introduction}

  Multi-epoch VLBI observations of BL~Lac object 0735+178 ($z=0.424$, 
Carswell et al. \cite{carswell 74}) by B\"a\"ath \& Zhang (\cite{BZ91}), 
B\"a\"ath et al. (\cite{Bal91}), Zhang \& B\"a\"ath (\cite{ZB91}), 
Gabuzda et al. (\cite{De94}) and G\'omez et al. (\cite{JL99}, \cite{JL01}) 
have shown the existence of superluminal motions with apparent velocities 
in the range of $\sim$ 6.5 to 12.2 $h^{-1}_{65}$ $c$ ($H_{\circ}$= 65 
$h_{65}$ km s$^{-1}$ Mpc$^{-1}$, $q_{\circ}$= 0.5). First indications of 
a strongly twisted curvature in 0735+178 were presented by Kellermann et al. 
(\cite{Ke98}) as part of a VLBI survey at 15 GHz.

  First polarimetric VLBI observations of this source were obtained by
Gabuzda, Wardle \& Roberts (\cite{De89}), showing a magnetic field
predominantly perpendicular to the jet axis. Polarimetric VLBI observations at
22 and 43 GHz by G\'omez et al. (\cite{JL99}) revealed a twisted jet with two
sharp apparent bends of 90$^{\circ}$ within two milliarcseconds of the
core. The magnetic field appears to smoothly follow one of the bends in the
jet, interpreted as perhaps produced by a precessing nozzle in the jet of
0735+178.

\begin{figure}
  \centerline{\psfig{figure=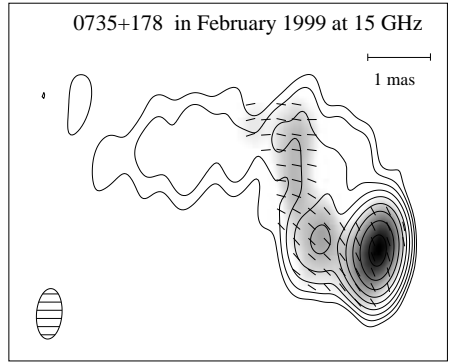,width=6cm}}
  \caption{15 GHz VLBI image of 0735+178 for epoch 27 February 1999. 
Contour levels represent total intensity for 0.3, 0.6, 1.2, 2.4, 4.8, 8.6, 
19.2,38.4,and 76.8\% of the peak intensity of 0.438 Jy/beam. Grey scale 
represents the polarized flux (peak at 5 mJy/beam with the noise at the 
level of 1 mJy/beam) and bars indicate the direction of the projected 
magnetic field vector. Convolving beam (shown as an ellipse) is 0.80x0.40 
mas with position angle of -5$^{\circ}$.}
\label{map}
\end{figure}

\section{The twisted jet structure of 0735+178}

  Figure \ref{map} shows a 15 GHz image obtained on 27 February 1999, as
part of a monitoring of 0735+178 consisting of 8, 15, 22 and 43 GHz VLBI
observations covering a total of 12 epochs, from March 1996 to February
2000. In agreement with previous observations (G\'omez et al. \cite{JL99}),
both total and polarized intensity images show two bends of 90 degrees
in the inner jet structure, with a magnetic field vector that follows the 
projected jet axis along the bends.

\begin{figure}
  \centerline{\psfig{figure=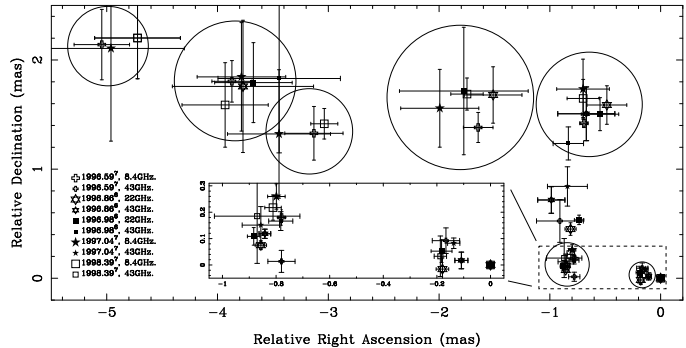,width=9cm}}
  \caption{Components position, relative to the core, in the jet of 0735+178
detected between 1996.59 and 1998.39. Epoch, frequency, reference, and symbol 
for each epoch are as labeled. Estimated errors are indicated by bars. 
References correspond to: {\it 6} G\'omez et al. (\cite{JL99}) and {\it 7} 
G\'omez et al. (\cite{JL01}.)}
\label{posxy}
\end{figure}

  To explain the sharp apparent structure in 0735+178 we need 
to assume a very small viewing angle for 0735+178, which would enhance the 
intrinsic curvature by projection effects. Figure \ref{posxy} summarizes the 
detected modelfit components in 0735+178
from 1996 to 1998 (from G\'omez et al. \cite{JL01}), restricted to the inner
5 milliarcseconds from the core. After 1996.59 all the observed components lye
in a well defined twisted structure coincident with our map presented in
Fig. \ref{map}. 
We could interpret the bending 
as produced either by a true change in the direction of ejection 
(e.g., jet precession) or by gradients in the external pressure not aligned 
with the initial direction of the jet flow. Precession of the jet allows, a 
priori, for motions in which the flow velocity vectors follow the jet bending 
(non-ballistic motion), or ballistic motions. In the next subsections we 
discuss these two different interpretations.

\subsection{Non Ballistic Superluminal Motion Model}

\begin{figure}
  \centerline{\psfig{figure=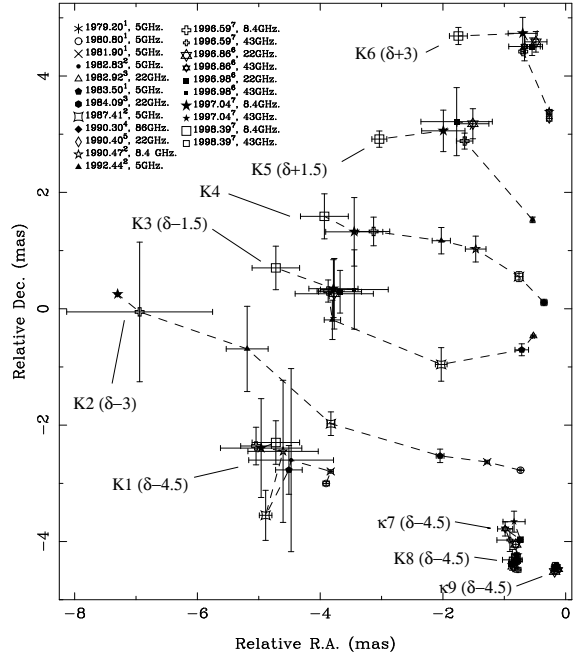,width=7.5cm}}
  \caption{Apparent trajectories of components in 0735+178. 
Components positions have been shifted in declination by 
the amount (in milliarcseconds) expressed in parenthesis. Epoch, frequency, 
reference, and symbol for each epoch are as labeled. Estimated errors are 
indicated by bars. References correspond to: {\it 1} B\"a\"ath \& Zhang 
(\cite{BZ91}); {\it 2} Gabuzda et al. (\cite{De94}); {\it 3} Zhang \& 
B\"a\"ath (\cite{ZB91}); {\it 4} Rantakyr\"o et al. (\cite{Ra98}); {\it 5} 
B\"a\"ath, Zhang, \& Chu (\cite{Bal91}); {\it 6} G\'omez et al. (\cite{JL99}) 
and {\it 7} G\'omez et al. (\cite{JL01}).}
\label{compev}
\end{figure}

  The observed magnetic vector parallel to the curved jet axis of 0735+178 
(Fig. \ref{map}) supports the idea of non-ballistic fluid motions. Further evidence 
of non-ballistic motions are obtained by analyzing the trajectories 
followed by the different components, shown in Fig. \ref{compev}, in 
which we observe that components K5 and K6 seem to move following a 
non-ballistic trajectory with superluminal motions of 10 $\pm$ 3.5 and 4.6 
$\pm$ 0.1 $h_{65}^{-1}$ c, respectively. However, we should note that this 
is based on uncertain identification of components through epochs,
in some cases more than 4 years apart.

  Figure \ref{compev} also reveals quasi-stationary components (K1, K8 and
k9), as well as some other components that seem to move following rectilinear
(ballistic) trajectories. That is the case for components K2, K3 and K4, moving
at apparent superluminal speeds of 11.6 $\pm$ 0.6, 8 $\pm$ 1.5, and 5 $\pm$ 1 
$h_{65}^{-1}$ c, respectively. There are some evidence of a possible change 
in the jet structure of 0735+178 (G\'omez et al. \cite{JL01}) suggesting that 
perhaps this rectilinear trajectory was actually present when these components
were detected. Therefore the apparent ballistic motions of these components 
may just be due to this initial rectilinear structure of the jet.

  The existence of standing and moving features is in agreement with
non-ballistic components, where the stationary components would be associated
with bends in the jet trajectory. Our observations support this scenario,
where component K8 is observed to remain quasi-stationary, and is located at
the first sharp bend of 90$^{\circ}$. A region of higher emission is also
observed in the second bent of 90$^{\circ}$, in which the jet turns back to its
initial east direction. Therefore, all these evidence suggest that
components in 0735+178 are not ballistic, following the local direction of the
jet flow, but they are not conclusive.

\subsection{Low Velocity ``Ballistic'' Motion Model}

  As pointed out in Fig. \ref{posxy}, components show a tendency to be 
clustered near the same position (indicated by circles), within the errors. 
Therefore, the actual 
structure of 0735+178 is also consistent with a quasi stationary situation 
in which all components would have remained at similar locations. 

\begin{figure}
  \centerline{\psfig{figure=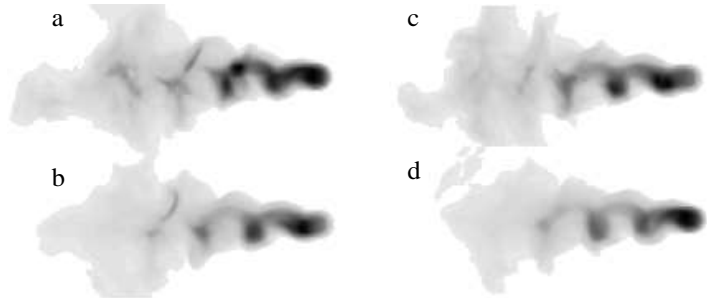,width=9.5cm}}  
  \caption{Synthetic synchrotron emission, for a viewing angle of $10^{\circ}$,
of a 3D hydrodynamical relativistic simulation for a jet with a helical  
perturbation in velocity. The four epochs, separated by the same amount of 
time, are labeled accordingly to the time-evolution.}
\label{sim3d}
\end{figure}

  To test this model we have performed three dimensional hydrodynamic
relativistic simulations of a precessing jet, using the numerical code
developed by Aloy et al. (\cite{AL00}, and references therein). Radio emission
from these HD results can be obtained by solving the transfer equations for
synchrotron radiation (G\'omez et al. \cite{JL97}). Figure \ref{sim3d} shows 
the resulting simulated total intensity maps for a viewing angle of 
$10^{\circ}$ and at different time steps. A twisted jet structure, similar 
to the observed in 0735+178, can easily be distinguished. Differential 
Doppler boosting result in the appearance of emission components in the jet 
regions that are better oriented with the observer. Because of the precession 
of the helix, these components seem to move following ballistic trajectories, 
even though the jet fluid moves along the jet axis, that is, with 
non-ballistic trajectories. It is possible that some of the components that 
we observe moving ballistically may be associated with the components 
produced by the differential Doppler boosting. On the other hand, the 
components observed to follow bent trajectories may be associated with moving 
shocks, that should follow a non-ballistic motion along the jet. In this 
scenario, ballistic components should move with a smaller velocity than the 
non-ballistic ones.

\section{Conclusions}

  Our new map of 0735+178 in 1999 shows a twisted geometry in the inner
jet regions as seen previously (G\'omez et al \cite {JL99}), but the 
observational evidence is not conclusive as to whether the twisted geometry 
of 0735+178 is produced by ballistic or non-ballistic motions of components. 
 Further observations, and their comparison with numerical simulations, 
should provide the necessary information to understand this unussual bent 
structure observed in 0735+178.

\end{document}